\documentclass[12pt, twoside]{article}

\usepackage{PRIMEarxiv}

\usepackage[utf8]{inputenc} 
\usepackage[T1]{fontenc}    
\usepackage{hyperref}       
\usepackage{url}            
\usepackage{booktabs}       
\usepackage{amsfonts}       
\usepackage{nicefrac}       
\usepackage{microtype}      
\usepackage{lipsum}
\usepackage{fancyhdr}       
\usepackage{graphicx}       
\graphicspath{{media/}}     

\usepackage{amsmath,amsfonts,amssymb}
\usepackage{multirow}
\usepackage{cite}

\title{Deep Learning Architectures for Medical Image Denoising: A Comparative Study of CNN-DAE, CADTra, and DCMIEDNet
}

\author{
  Asadullah Bin Rahman, Masud Ibn Afjal, Md. Abdulla Al Mamun\\
  Department of Computer Science and Engineering\\
  Hajee Mohammad Danesh Science and Technology University (HSTU)\\
  Dinajpur -- 5200, Bangladesh\\
  \texttt{galib.cse.17020221@std.hstu.ac.bd, {\{masud, mamun\}@hstu.ac.bd}} \\
}

\begin{document}
\maketitle

\begin{abstract}
Medical imaging modalities are inherently susceptible to noise contamination that degrades diagnostic utility and clinical assessment accuracy. This paper presents a comprehensive comparative evaluation of three state-of-the-art deep learning architectures for MRI brain image denoising: CNN-DAE, CADTra, and DCMIEDNet. We systematically evaluate these models across multiple Gaussian noise intensities ($\sigma = 10, 15, 25$) using the Figshare MRI Brain Dataset. Our experimental results demonstrate that DCMIEDNet achieves superior performance at lower noise levels, with PSNR values of $32.921 \pm 2.350$ dB and $30.943 \pm 2.339$ dB for $\sigma = 10$ and $15$ respectively. However, CADTra exhibits greater robustness under severe noise conditions ($\sigma = 25$), achieving the highest PSNR of $27.671 \pm 2.091$ dB. All deep learning approaches significantly outperform traditional wavelet-based methods, with improvements ranging from 5-8 dB across tested conditions. This study establishes quantitative benchmarks for medical image denoising and provides insights into architecture-specific strengths for varying noise intensities.

\end{abstract}

\section{Introduction}

Medical imaging modalities—X-ray, MRI, CT, and ultrasound—are fundamental to modern diagnostics and treatment planning. However, these techniques are inherently susceptible to various noise sources that manifest as granular textures, blurring, or artifacts, obscuring fine anatomical details crucial for accurate clinical assessment. Noise origins include the physics of image acquisition, detector imperfections, and deliberate dose reduction to minimize patient radiation exposure, where lower doses correlate with increased noise levels.

Noise significantly degrades diagnostic utility, potentially leading to diagnostic errors, delayed treatment, or repeated imaging procedures, thereby increasing patient burden and healthcare costs. Consequently, effective image denoising serves as an indispensable preprocessing step in medical image analysis pipelines, aiming to suppress noise while preserving essential structural information and edge details.

Traditional wavelet-based methods have historically provided solid foundations for image denoising through multi-resolution analysis and frequency domain separation. However, these approaches often struggle with complex noise patterns and may introduce artifacts or over-smoothing effects.

Deep learning techniques, particularly convolutional neural networks (CNNs), have emerged as leading approaches due to their ability to learn complex mappings between noisy and clean images directly from data. CNNs leverage hierarchical feature extraction through convolutional layers \cite{majeed_zangana_classical_2024}. Various architectures have been proposed, including residual learning networks (ResNet) that learn noise residuals rather than clean images directly, and techniques like batch normalization to improve training stability \cite{izadi_image_2023, jifara_medical_2019, liu_densely_2020}.

Other deep learning architectures applied to denoising include Recurrent Neural Networks (RNNs), Variational Autoencoders (VAEs), and Transformer-based models \cite{atal_optimal_2023}. Denoising autoencoders learn robust feature representations for noise removal \cite{gondara_medical_2016, el-shafai_efficient_2022}. Generative Adversarial Networks (GANs) generate realistic denoised images, with variants like Enhanced SRGAN showing promise in low-dose PET denoising \cite{kaur_complete_2023}. Transformer-based models, such as Swin Transformer architectures like StruNet, adapt to different noise types across modalities including CT, OCT, and OCTA \cite{kaur_complete_2023, ma_strunet_2023}.

Deep learning models demonstrate state-of-the-art performance, maintaining details and producing sharper images compared to traditional methods. They effectively handle complex real noise and image restoration problems, including artifact reduction in medical images \cite{kaur_complete_2023}. However, challenges remain in developing unified frameworks capable of handling noise complexity across different medical imaging modalities, requiring vast training data and addressing overfitting with limited datasets.

This research implements and evaluates three deep learning architectures for MRI brain image denoising, providing comprehensive performance analysis and establishing benchmarks for future development.

Key contributions include:

\begin{itemize}
    \item Comprehensive performance benchmarking of three deep learning architectures across multiple noise levels
    \item Demonstration of architecture-specific strengths: DCMIEDNet for low-to-moderate noise and CADTra for high noise scenarios
    \item Establishment of quantitative baselines for future deep learning-based medical image denoising research
\end{itemize}

\section{Methodology}

Three relevant deep learning denoising techniques were implemented for MRI images across various noise settings, enabling comprehensive comparative analysis.

\subsection{Deep Learning Architectures}

\subsubsection{CNN-DAE Model}
The CNN-DAE method \cite{gondara_medical_2016}, developed by Gondara, addresses the perception that deep learning models require vast training data for optimal performance—a challenge in medical domains due to data scarcity and privacy concerns. The hypothesis centers on using convolutional denoising autoencoders for efficient medical image denoising with small sample sizes.

The architecture employs a symmetric encoder-decoder structure with convolutional layers, detailed in Table \ref{tab:cnn_dae_arch}.

\begin{table*}[!htb]
\centering
\caption{CNN-DAE Model Architecture Specifications}
\label{tab:cnn_dae_arch}
\begin{tabular}{l|l|c|c|c}
\toprule
\textbf{Stage} & \textbf{Layer Name} & \textbf{Operation} & \textbf{Output Shape} & \textbf{Parameters} \\
\midrule
\multirow{6}{*}{\textbf{Encoder}} 
& \texttt{input\_layer} & Input (grayscale) & $(224, 224, 1)$ & 0 \\
& \texttt{conv2d\_1} & Conv2D + ReLU & $(224, 224, 32)$ & 320 \\
& \texttt{max\_pool\_1} & MaxPooling2D ($2 \times 2$) & $(112, 112, 32)$ & 0 \\
& \texttt{conv2d\_2} & Conv2D + ReLU & $(112, 112, 64)$ & 18,496 \\
& \texttt{max\_pool\_2} & MaxPooling2D ($2 \times 2$) & $(56, 56, 64)$ & 0 \\
& \texttt{conv2d\_3} & Conv2D + ReLU & $(56, 56, 64)$ & 36,928 \\
\midrule
\multirow{4}{*}{\textbf{Decoder}}
& \texttt{up\_sample\_1} & UpSampling2D ($2 \times 2$) & $(112, 112, 64)$ & 0 \\
& \texttt{conv2d\_4} & Conv2D + ReLU & $(112, 112, 32)$ & 18,464 \\
& \texttt{up\_sample\_2} & UpSampling2D ($2 \times 2$) & $(224, 224, 32)$ & 0 \\
& \texttt{conv2d\_5} & Conv2D + Sigmoid & $(224, 224, 1)$ & 289 \\
\midrule
\multicolumn{4}{r|}{\textbf{Total Trainable Parameters}} & \textbf{74,497} \\
\bottomrule
\end{tabular}
\end{table*}

\subsubsection{CADTra Model}
The CADTra (Classification, Autoencoder Denoising, Transfer learning) model \cite{el-shafai_efficient_2022}, introduced by El-Shafai et al., is an automated system for efficient pneumonia-related disease diagnosis, including COVID-19, from chest X-rays and CT scans. The Autoencoder Denoising component serves as a dedicated preprocessing stage, mitigating effects of Gaussian, salt and pepper, and speckle noise commonly encountered in medical imaging.

The architecture consists of an eight-layer convolutional autoencoder network, detailed in Table \ref{tab:cadtra_arch}.

\begin{table*}[!htb]
\centering
\caption{CADTra Model Architecture Specifications}
\label{tab:cadtra_arch}
\begin{tabular}{l|l|l|c|c}
\toprule
\textbf{Stage} & \textbf{Layer Name} & \textbf{Operation} & \textbf{Output Shape} & \textbf{Parameters} \\
\midrule
\multirow{5}{*}{\textbf{Encoder}}
& \texttt{input\_layer} & Input (grayscale) & $(224, 224, 1)$ & 0 \\
& \texttt{batch\_norm\_1} & BatchNormalization & $(224, 224, 1)$ & 4 \\
& \texttt{conv2d\_1} & Conv2D + ReLU (128 filters, $3 \times 3$) & $(224, 224, 128)$ & 1,280 \\
& \texttt{conv2d\_2} & Conv2D + ReLU (64 filters, $3 \times 3$) & $(224, 224, 64)$ & 73,792 \\
& \texttt{conv2d\_3} & Conv2D + ReLU (32 filters, $3 \times 3$) & $(224, 224, 32)$ & 18,464 \\
\midrule
\multirow{4}{*}{\textbf{Decoder}}
& \texttt{conv2d\_trans\_1} & Conv2DTranspose + ReLU (32 filters, $3 \times 3$) & $(224, 224, 32)$ & 9,248 \\
& \texttt{conv2d\_trans\_2} & Conv2DTranspose + ReLU (64 filters, $3 \times 3$) & $(224, 224, 64)$ & 18,496 \\
& \texttt{conv2d\_trans\_3} & Conv2DTranspose + ReLU (128 filters, $3 \times 3$) & $(224, 224, 128)$ & 73,856 \\
& \texttt{conv2d\_output} & Conv2D + Sigmoid (1 filter, $3 \times 3$) & $(224, 224, 1)$ & 1,153 \\
\midrule
\multicolumn{4}{r|}{\textbf{Total Trainable Parameters}} & \textbf{196,293} \\
\bottomrule
\end{tabular}
\end{table*}

\subsubsection{DCMIEDNet Model}
The Dual Convolutional Medical Image-Enhanced Denoising Network (DCMIEDNet) \cite{sahu_application_2023}, proposed by Sahu et al., is designed for chest X-ray image denoising, particularly for Additive White Gaussian Noise (AWGN). The architecture derives from the DudeNet model, adapted for medical imaging applications.

DCMIEDNet comprises four main components: Feature Extraction Block (FEB), Enhancement Block (EB), Compression Block (CB), and Reconstruction Block (RB). The FEB includes two parallel subnetworks: SubNet1 uses sparse mechanisms combining standard and dilated convolutions across 16 layers, while SubNet2 employs a simpler 16-layer stack ending in 1×1 compression. The model leverages residual learning, multi-scale feature extraction, and efficient compression strategies, totaling approximately 1.49 million parameters (Table \ref{tab:dcmiednet_arch}, \ref{tab:dcmiednet_detailed}).

\begin{table*}[!htb]
\centering
\caption{DCMIEDNet Model Architecture Overview and Component Specifications}
\label{tab:dcmiednet_arch}
\begin{tabular}{l|l|c|c}
\toprule
\textbf{Component} & \textbf{Description} & \textbf{Key Operations} & \textbf{Parameters} \\
\midrule
\multirow{3}{*}{\textbf{SubNet1}} 
& Feature Extraction Block (FEB) & Conv + BN + ReLU (Layers 1,3,4,6-8,10,11,13-15) & \multirow{3}{*}{$\sim$745K} \\
& Sparse Processing & Dilated Conv + BN + ReLU (Layers 2,5,9,12) & \\
& Output Layer & Conv2D (Layer 16) & \\
\midrule
\multirow{2}{*}{\textbf{SubNet2}}
& Standard Processing & Conv + ReLU (Layers 1-15) & \multirow{2}{*}{$\sim$372K} \\
& Compression Block [I] & $1 \times 1$ Conv (Layer 16) & \\
\midrule
\textbf{Fusion Stage} & Feature Concatenation & Concat(SubNet1, SubNet2) & 0 \\
\midrule
\multirow{4}{*}{\textbf{Post-Fusion}}
& Enhancement Block [I] & Multi-scale feature enhancement & \multirow{4}{*}{$\sim$376K} \\
& Compression Block [II] & Dimensionality reduction & \\
& Enhancement Block [II] & Secondary enhancement & \\
& Compression Block [III] & Final compression & \\
\midrule
\textbf{Reconstruction} & Residual Learning Block & Noise estimation + subtraction & $\sim$1K \\
\midrule
\multicolumn{3}{r|}{\textbf{Total Trainable Parameters}} & \textbf{1,493,024} \\
\bottomrule
\end{tabular}
\end{table*}

\begin{table*}[!htb]
\centering
\caption{DCMIEDNet Detailed SubNetwork Layer Specifications}
\label{tab:dcmiednet_detailed}
\begin{tabular}{c|l|l|c|c}
\toprule
\textbf{SubNet} & \textbf{Layer Range} & \textbf{Operation Type} & \textbf{Filter Size} & \textbf{Activation} \\
\midrule
\multirow{3}{*}{\textbf{SubNet1}}
& Layers 1,3,4,6-8,10,11,13-15 & Standard Conv + BN + ReLU & $3 \times 3$ & ReLU \\
& Layers 2,5,9,12 & Dilated Conv + BN + ReLU & $3 \times 3$ (dilation=2) & ReLU \\
& Layer 16 & Conv2D (output) & $3 \times 3$ & Linear \\
\midrule
\multirow{2}{*}{\textbf{SubNet2}}
& Layers 1-15 & Standard Conv + ReLU & $3 \times 3$ & ReLU \\
& Layer 16 (CB[I]) & Compression Conv & $1 \times 1$ & Linear \\
\midrule
\multirow{4}{*}{\textbf{Post-Fusion}}
& EB[I] & Enhancement Block & Multi-scale & ReLU \\
& CB[II] & Compression Block & $1 \times 1$ & Linear \\
& EB[II] & Enhancement Block & Multi-scale & ReLU \\
& CB[III] & Compression Block & $1 \times 1$ & Linear \\
\midrule
\textbf{Reconstruction} & RB & Residual Block & $3 \times 3$ & Linear \\
\bottomrule
\end{tabular}
\end{table*}

\subsection{Experimental Setup}

Experiments were conducted in Google Colab using T4 GPU acceleration for deep neural network training. The objective focused on removing Gaussian noise from brain MRI scans using standardized parameters: $224 \times 224$ pixel image resizing and consistent noise profiles with mean 0 and normalized standard deviations of 10, 15, and 25.

Implementation utilized two major frameworks: TensorFlow/Keras for CNN-DAE and CADTra models, and PyTorch for DCMIEDNet.

\subsubsection{Data Preparation}
A uniform pipeline ensured fair model evaluation. The Figshare MRI Brain Dataset \cite{jun_cheng_brain_2017} was applied, with each image undergoing two-step preprocessing: pixel value normalization to 0-1 range and resizing to $224 \times 224$ dimensions. These constituted clean ground truth data. Gaussian noise was artificially added to create noisy inputs (Table \ref{tab:noise-configDEEP1}).

Dataset splitting allocated 80\% for training and 20\% for testing, with 15\% of training data reserved for validation. Random state 42 ensured reproducibility.

\begin{table}[!htb]
\centering
\caption{Noise Configuration in Experiments}
\label{tab:noise-configDEEP1}
\begin{tabular}{ccc}
\toprule
\begin{tabular}[c]{@{}c@{}}Noise Standard \\ Deviation ($\sigma$)\end{tabular} 
& SSIM & PSNR \\
\midrule
10 & 0.500 & 19.316 \\
15 & 0.385 & 17.531 \\
25 & 0.252 & 14.128 \\
\bottomrule
\end{tabular}
\end{table}

\subsubsection{Training Configuration}
Mean Squared Error (MSE) loss function guided training, quantifying pixel-wise differences between denoised outputs and clean images. Adam optimizer minimized this error across all models.

CNN-DAE and CADTra models (TensorFlow/Keras) trained for 100 epochs with batch size 5, regulated by Early Stopping callback (5-epoch patience). DCMIEDNet (PyTorch) followed similar parameters but employed continuous validation loss tracking, saving optimal model states. Training hyperparameters are summarized in Table \ref{tab:hyperparameters1}.

\begin{table}[h]
\centering
\caption{Model Hyperparameters}
\label{tab:hyperparameters1}
\begin{tabular}{ll}
\toprule
\textbf{Hyperparameter} & \textbf{Value} \\
\midrule
Image Size     & $224 \times 224$ \\
Loss Function  & Mean Squared Error (MSE) \\
Optimizer      & Adam \\
Epochs         & 100 \\
Batch Size     & 5 \\
Learning Rate  & 0.001 \\
\bottomrule
\end{tabular}
\end{table}

\section{Results}

Comprehensive evaluation across three noise levels revealed distinct performance characteristics for each deep learning architecture (Table \ref{tab:dl_summary}).

\begin{table}[!htb]
\centering
\caption{Deep Learning-Based Denoising Performance Summary}
\label{tab:dl_summary}
\begin{tabular}{c|l|c|c}
\toprule
\textbf{$\sigma$} & \textbf{Method} & \textbf{PSNR (dB)} & \textbf{SSIM} \\
\midrule
\multirow{3}{*}{10} & CADTra & $31.895 \pm 2.431$ & $0.847 \pm 0.061$ \\
& CNN-DAE & $29.972 \pm 1.764$ & $0.847 \pm 0.047$ \\
& DCMIEDNet & $32.921 \pm 2.350$ & $0.823 \pm 0.068$ \\
\midrule
\multirow{3}{*}{15} & CADTra & $29.187 \pm 2.410$ & $0.805 \pm 0.052$ \\
& CNN-DAE & $28.616 \pm 1.798$ & $0.817 \pm 0.040$ \\
& DCMIEDNet & $30.943 \pm 2.339$ & $0.796 \pm 0.065$ \\
\midrule
\multirow{3}{*}{25} & CADTra & $27.671 \pm 2.091$ & $0.766 \pm 0.062$ \\
& CNN-DAE & $26.575 \pm 1.834$ & $0.750 \pm 0.064$ \\
& DCMIEDNet & $27.081 \pm 2.570$ & $0.715 \pm 0.080$ \\
\bottomrule
\end{tabular}
\end{table}

\textbf{Low Noise Level ($\sigma = 10$)}: All models demonstrated strong performance. DCMIEDNet achieved the highest PSNR of $32.921 \pm 2.350$ dB, followed by CADTra ($31.895 \pm 2.431$ dB) with superior SSIM ($0.847 \pm 0.061$). CNN-DAE yielded competitive results with PSNR of $29.972 \pm 1.764$ dB.

\textbf{Moderate Noise Level ($\sigma = 15$)}: Performance decreased correspondingly across all models while maintaining robustness. DCMIEDNet continued leading with PSNR of $30.943 \pm 2.339$ dB. CADTra and CNN-DAE followed with PSNRs of $29.187 \pm 2.410$ and $28.616 \pm 1.798$ dB, respectively, with CNN-DAE showing the highest SSIM ($0.817 \pm 0.040$).

\textbf{High Noise Level ($\sigma = 25$)}: The most challenging scenario revealed interesting performance shifts. CADTra emerged as the optimal model with both highest PSNR ($27.671 \pm 2.091$ dB) and SSIM ($0.766 \pm 0.062$), followed by DCMIEDNet ($27.081 \pm 2.570$ dB) and CNN-DAE ($26.575 \pm 1.834$ dB).

\section{Discussion}

The experimental results demonstrate the effectiveness of deep convolutional neural networks for MRI denoising across varying noise levels. The comparative analysis reveals several key insights:

\textbf{Architecture Performance}: DCMIEDNet's superior performance at lower noise levels ($\sigma = 10, 15$) can be attributed to its sophisticated dual-path architecture with multi-scale feature extraction and residual learning mechanisms. However, CADTra's emergence as the optimal model at high noise levels ($\sigma = 25$) suggests that its deeper encoder-decoder structure with batch normalization provides better robustness against severe noise contamination.

\textbf{Model Complexity vs. Performance}: The relationship between model complexity and performance is not strictly linear. While DCMIEDNet has significantly more parameters (1.49M) than CADTra (196K) and CNN-DAE (74K), its advantage diminishes at higher noise levels, potentially indicating overfitting or sensitivity to severe noise conditions.

\textbf{Framework Implementation}: The successful implementation across TensorFlow/Keras and PyTorch frameworks demonstrates the reproducibility and generalizability of the approaches, though direct performance comparisons must account for potential framework-specific optimizations.

\textbf{Limitations and Future Directions}: The study's scope is limited to synthetic Gaussian noise, while real-world MRI noise typically follows Rician distributions. The models were evaluated on a single dataset from Figshare, potentially limiting generalizability across different scanners, protocols, and patient demographics. Future work will incorporate more realistic noise models, diverse datasets, and advanced architectures such as U-Nets or Generative Adversarial Networks.

\section{Conclusion}

This research establishes that deep convolutional neural networks provide powerful and viable solutions for MRI denoising. The systematic evaluation of three distinct architectures—CNN-DAE, CADTra, and DCMIEDNet—demonstrates their effectiveness in learning complex mappings for additive Gaussian noise removal while preserving anatomical structures.

The significance lies in the potential to enhance diagnostic imaging quality through automated denoising, potentially improving clinical assessment accuracy and reliability. While findings are promising, clinical deployment requires rigorous validation on real-world data with complex noise characteristics. This work provides a solid foundation demonstrating deep learning's substantial advantages over traditional filtering methods, establishing clear pathways for advancing medical imaging technology.

\bibliographystyle{ieeetr}
\nocite{*}
\bibliography{main}

\end{document}